\documentclass{article}
\usepackage{spconf,amsmath,graphicx}
\usepackage{color}
\usepackage[table]{xcolor}

\title{\vspace{-12mm}On loss functions and evaluation metrics for music source separation}
\name{Enric Gusó$^{\star}$, Jordi Pons$^{\dagger}$, Santiago Pascual$^{\dagger}$, Joan Serrà$^{\dagger}$\vspace{-2mm}}
\address{Music Technology Group, Universitat Pompeu Fabra$^{\star}$ \hspace{8mm} Dolby Laboratories$^{\dagger}$\vspace{-4mm}}
\begin{document}
\ninept
\maketitle

\begin{abstract}
We investigate which loss functions provide better separations via benchmarking {an extensive set of those for music source separation.}
To that end, we first survey the most representative audio source separation losses we identified, to later consistently benchmark them in a controlled experimental setup. {We also explore using such losses as evaluation metrics, via cross-correlating them with the results of a subjective test.} Based on the observation that the standard signal-to-distortion ratio metric can be misleading in some scenarios, we study {alternative evaluation metrics based on the considered losses.} 

\end{abstract}
\vspace{-2mm}

\section{Introduction}\label{sec:introduction}
\vspace{-1mm}

Sound mixing consists in bringing together different audio signals (sources) to create a mix. Source separation is the inverse process of un-mixing, with the goal of recovering the original signals. Our work focuses on music source separation,
and on which training objectives (losses) provide separations that are perceptually preferable. To that end, we benchmark an extensive set of regression losses that are representative of the recent literature on audio source separation (Fig.~1).
Regression losses are also often used as evaluation metrics, since they are a quick alternative to subjective tests~\cite{Roux_2019,kadiouglu2020empirical,liu2021permutation}. However, these can be problematic~\cite{cano2016evaluation}.
Thus, we also study their appropriateness for evaluation by correlating them with human judgment.

Many successful music source separation methods rely on regressors operating on the time-frequency domain.
These can be used to estimate the sources' magnitude spectrograms~\cite{stoter2019open,chandna2017monoaural}, or to estimate separation masks to filter the magnitude spectrogram of the mixture. 
While spectrogram-based models are widely used~\cite{stoter2019open,chandna2017monoaural,roebel2015automatic}, 
these can be problematic as they often discard the phase, which is needed to go back to the time domain.
To solve that, the phase of the mixture is often used~\cite{stoter2019open,chandna2017monoaural,roebel2015automatic}, but this can introduce errors. Our work considers the use of phase-sensitive and time-domain losses to capture phase errors within the training objective.
Further, note that scale-invariant time-domain losses are now becoming popular for audio source separation~\cite{Roux_2019,liu2021permutation}. Our work also investigates the use of scale-invariant regression losses for spectrogram signals~\cite{fevotte2009nonnegative}.
Finally, if we look at training losses from a machine learning perspective, it is known that
minimizing the $L1$ or $L2$ loss leads to predicting the median or the  mean~\cite{zen2014deep}, respectively, of the output distribution that is assumed to be uni-modal. Consequently, if the underlying distribution is not as assumed, predictions may not follow the true distribution, introducing errors.
This is commonly tackled with regularizers: adversarial losses that help fitting a ``realistic'' mode, and deep feature losses that provide additional cues during training.

Our main contributions are: \textit{{i)}} a survey of the most representative losses for neural audio source separation, in section 2; \textit{{ii)}} to benchmark such losses, in sections 3 and 4; and 
\textit{{iii)}} a subjective evaluation to study which losses correlate best with human judgments, in section~5. 
Note that to consistently benchmark the considered losses one needs to choose a specific model. We select OpenUnmix~\cite{stoter2019open} because it is an open-source\footnote{{https://github.com/sigsep/open-unmix-pytorch}} state-of-the-art model based on predicting time-frequency masks, what is
significantly faster to train than waveform-based models~\cite{defossez2019music,pons2020upsampling}.

\section{Audio source separation losses}

\begin{figure}[t!]
	\label{fig:taxonomy}
	\centering
	\includegraphics[clip,width=0.90\columnwidth]{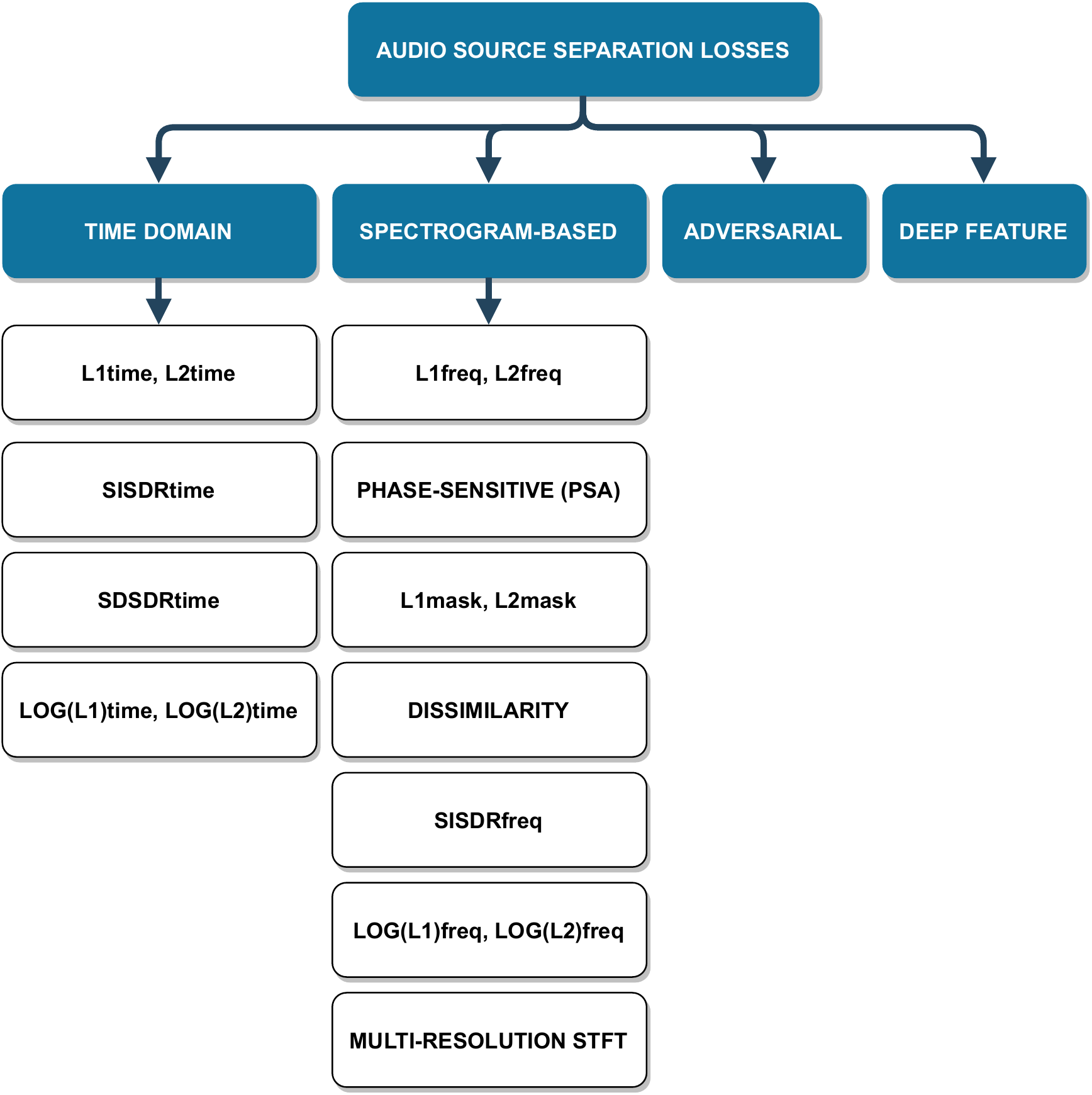}%
	\vspace{-2mm}
	\caption{Taxonomy of the considered audio source separation losses.}
\end{figure}

\subsection{Time domain losses}\label{sec:timelosses}

Given an audio $x_t$ of {\Large$\tau$} samples, we extract $K$ different sources $\tilde{y}_{t,k}$.
Standard regression losses, like L1 or L2, have been used to train audio source separation models~\cite{stoter2019open,chandna2017monoaural}:

\begin{gather}
\label{l1_time}
\text{L1}_\text{time} = \frac{1}{\scalebox{1.5}{$\tau$}K} \sum_{t , k}|\tilde{y}_{t, k}-y_{t, k}| , \notag\\
\text{L2}_\text{time} = \frac{1}{\scalebox{1.5}{$\tau$}K} \sum_{t, k}|\tilde{y}_{t, k}
-y_{t, k}|^{2} . \notag
\end{gather}
Further, signal-to-distortion rartio (SDR)~\cite{vincent2006performance} and scale-invariant SDR (SISDR)~\cite{Roux_2019} are widely used for evaluation and training~\cite{luo2019conv}:
\begin{gather}
\label{sisdr}
\text{SISDR}_\text{time}=\frac{-10}{\scalebox{1.5}{$\tau$}K} \sum_{t , k} \log_{10}\left(\frac{\left|\frac{\tilde{y}_{t,k}^{\text{T}}y_{t,k}}{|y_{t,k}|^{2}}y_{t,k}\right|^{2}}{\left|\frac{\tilde{y}_{t,k}^{T}y_{t,k}}{|y_{t,k}|^{2}}y_{t,k}-\tilde{y}_{t,k}\right|^{2}}\right) . \notag
\end{gather}
However, in some cases, the scale-invariance assumption might not hold and the scale-dependent SDR (SDSDR)~\cite{Roux_2019} is preferable:
\begin{gather}
\label{sdsdr}
\text{SNR} = \frac{10}{\scalebox{1.5}{$\tau$}K} \sum_{t , k} \log_{10}\left(\frac{|y_{t,k}|^{2}}{ |\tilde{y}_{t,k}-y_{t,k}|^{2}}\right) ,\notag\\
\text{L}_\text{down} = SNR + \frac{10}{\scalebox{1.5}{$\tau$}K} \sum_{t , k}\log_{10} \left(\left(\frac{\tilde{y}_{t,k}^{\text{T}}y_{t,k}}{|y_{t,k}|^{2}}\right)^{2}\right) ,\notag \\
\text{SDSDR}_\text{time} = {-}\min(\text{SNR}, \text{L}_\text{down}).\notag
\end{gather}
Further, Heitkaemper et al.~\cite{heitkaemper2019demystifying} derived a log-compressed L2 loss, equivalent to SISDR. We also experiment with an L1 variant:
\begin{gather}
 \label{eq:similarL2}
 \text{LOGL2}_\text{time}=\frac{10}{\scalebox{1.5}{$\tau$}K}\sum_{k}\log_{10}\sum_{t}|\tilde{y}_{t,k}-y_{t,k}|^{2} ,\notag
\\ \text{LOGL1}_\text{time}=\frac{10}{\scalebox{1.5}{$\tau$}K}\sum_{k}\log_{10}\sum_{t}|\tilde{y}_{t,k}-y_{t,k}|.\notag
\end{gather}

\subsection{Spectrogram-based losses}\label{subsec:speclosses}

We define the short-time Fourier transform as $STFT(x_t)=X_{n,\omega}$, where $n$ and $\omega$ stand for frame and frequency bin indices, respectively, of a total number of $N$ frames and $\Omega$ frequency bins. 
Many source separation models estimate a magnitude spectrogram $|\tilde{Y}_{n,\omega,k}|$ per source $k$, and rely on standard regression losses like L1 or L2:
\begin{gather}
\label{eq2}
\text{L1}_\text{freq} = \frac{1}{N\Omega K} \sum_{n,\omega,k}\left|\hspace{1mm}|\tilde{Y}_{n,\omega,k}|-|Y_{n,\omega,k}|\hspace{1mm}\right|,\notag\\
\text{L2}_\text{freq} = \frac{1}{N\Omega K} \sum_{n,\omega,k}\left|\hspace{1mm}|\tilde{Y}_{n,\omega,k}|
-|{Y}_{n,\omega,k}|\hspace{1mm}\right|^{2}.\notag
\end{gather}
Note that $|\tilde{Y}_{n,\omega,k}|$ does not include phase information. 
To synthesize the time-domain signal from $|\tilde{Y}_{n,\omega,k}|$ with the inverse short-time Fourier transform, phase information is required. 
Often times, the noisy phase of the input mixture $\angle X_{n,\omega}$ is used: $\tilde{y}_{t,k}=ISTFT(|\tilde{Y}_{n,\omega,k}|\angle X_{n,\omega})$, which can introduce errors. 
To mitigate those, a phase sensitive target $Y^\text{PSA}_{n,\omega,k}$ was proposed~\cite{erdogan2015phase}:
\begin{gather}
\label{eq:PSA}
Y^\text{PSA}_{n,\omega,k}=|Y_{n,\omega,k}|\cos(\angle X_{n,\omega} - \angle Y_{n,\omega,k}), \notag\\
\text{L}_\text{PSA}=\frac{1}{N\Omega K}\sum_{n,\omega,k}\left| \hspace{1mm}|\tilde{Y}_{n,\omega,k}|-Y^\text{PSA}_{n,\omega, k}\hspace{1mm}\right|^{2}.\notag
\end{gather}
Further, mask-based models are commonly used when operating with spectrogram signals and, e.g., ratio masks are widely employed~\cite{stoter2019open,tesisenric} to filter the sources out from the mixture $X_{n, \omega}$:
\begin{gather}
M_{n,\omega,k} = \frac{|Y_{n, \omega,k}|}{\sum_{k}|Y_{n,\omega,k}|},\notag \\
|{Y}_{n,\omega,k}| = M_{n,\omega,k} \odot |X_{n, \omega}| .\notag
\end{gather}
Standard regression losses are used to train mask-based models~\cite{stoter2019open}:
\begin{gather}
\text{L1}_\text{mask} = \frac{1}{N\Omega K} \sum_{n,\omega,k}|\tilde{M}_{n,\omega , k}-{M}_{n, \omega,k}| ,\notag \\
\text{L2}_\text{mask} = \frac{1}{N\Omega K} \sum_{n, \omega,k}|\tilde{M}_{n, \omega,k}
-M_{n, \omega,k}|^{2}.\notag
\end{gather}
The above losses encourage separations to be similar to the targets. Yet, source separation models are prone to introduce interferences from other sources. To mitigate this, Huang et al.~\cite{huang2014deep} proposed a dissimilarity loss encouraging each prediction to be different from the rest 
by adding an L2 term for each additional source~\cite{huang2014deep,chandna2017monoaural}:
\begin{gather}
\label{eq:dissimilarity}
\text{L}_\text{dissim}
=\frac{1}{N\Omega K}\sum_{n, \omega,k} \bigg( \left|\hspace{1mm}|\tilde{Y}_{n,\omega,k}|-|Y_{n,\omega,k}|\hspace{1mm}\right|^{2} - \notag \\ -\beta\sum_{\tilde{k}}\left|\hspace{1mm}|\tilde{Y}_{n,\omega,k}|-|Y_{n,\omega,\tilde{k}\neq k}|\hspace{1mm}\right| ^{2} \bigg).\notag
\end{gather}
We also adapt SISDR$_\text{time}$ to work with spectrograms, reshaping $Y_{n,k,\omega}$ and $\tilde{Y}_{n,k,\omega}$ into $Y_{n\omega,k}$ and $\tilde{Y}_{n\omega,k}$, respectively. Further, we adapt LOGL1$_\text{time}$ and LOGL2$_\text{time}$ for spectrograms:
\begin{gather}
\text{LOGL1}_\text{freq} = \frac{10}{N\Omega K} \sum_{k}\log_{10}\sum_{n,\omega}\left|\hspace{1mm}|\tilde{Y}_{n,\omega,k}|-|Y_{n, \omega,k}|\hspace{1mm}\right| ,\notag\\
\text{LOGL2}_\text{freq} = \frac{10}{N\Omega K} \sum_{k}\log_{10}\sum_{n,\omega}\left|\hspace{1mm}|\tilde{Y}_{n,\omega,k}|-|Y_{n,\omega,k}|\hspace{1mm}\right|^{2}.\notag
\end{gather}
We also study the multi-resolution STFT loss (L$_\text{MRS}$), 
implemented via 
L$_\text{sc}$ (spectral convergence) and L$_\text{mag}$ (log magnitude)~\cite{yamamoto2020parallel,arik2018fast}.
It is composed of single-resolution STFTs 
with different configurations $O$ having various FFT sizes, windows and overlaps:
\begin{gather}
\text{L}_\text{sc}=\frac{1}{K} \sum_{k}\frac{\sqrt{\sum_{n,\omega}(\hspace{1mm} |\tilde{Y}_{n,\omega,k}|-|Y_{n,\omega,k}|\hspace{1mm})^{2}}}{\sqrt{\sum_{n,\omega}|Y_{n,\omega,k}|^{2}}},\notag \\
\text{L}_\text{mag} = \frac{1}{K} \sum_{n,\omega,k}(\log_{10}|\tilde{Y}_{n,\omega,k}| - \log_{10}|{Y}_{n,\omega,k}|),\notag \\
\text{L}_\text{MRS}= \frac{1}{O}\sum_{o} \left( \text{L}_\text{sc}^{(o)} + \text{L}_\text{mag}^{(o)} \right) .\notag
\end{gather} 

\vspace{-7mm}
\subsection{Adversarial loss}
\label{sec:adversarial}

\begin{figure}[t!]
	\centering
	\includegraphics[trim={0 0 0 1.9cm},clip,width=0.8\columnwidth]{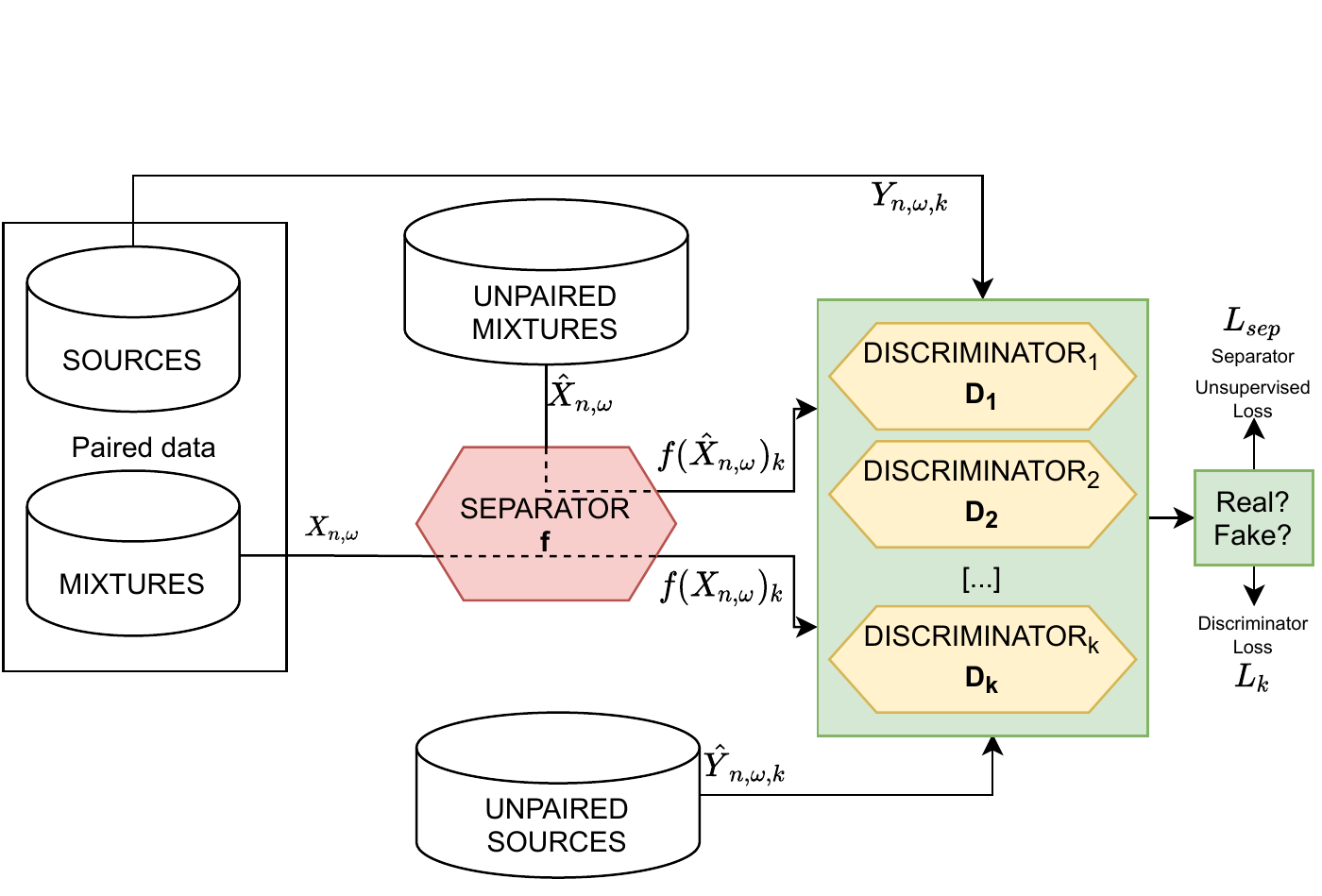}%
    \vspace{-2mm}
	\caption{Adversarial loss schema.
	}
	\label{fig:adv1}
		\vspace{-2mm}

\end{figure}

Our adversarial loss~\cite{goodfellow2020generative,pascual2017segan,stoller2018adversarial} (Fig.~\ref{fig:adv1}) consists of \textit{i)} a separator $f$ producing plausible separations $\tilde{Y}$ (magnitude spectrograms) from paired $X$ and unpaired $\hat{X}$ mixtures; and \textit{ii)} several discriminators~$D$, dictating whether they are produced by the separator $\tilde{Y}$ (fake) or come from a database of paired $Y$ or unpaired $\hat{Y}$ isolated recordings (real). 
As a result, this setup allows using additional (unpaired) data to train the separator (see in Fig.~\ref{fig:adv1}). 
We use a least-squares variant~\cite{mao2017least} with label smoothing~\cite{salimans2016improved,szegedy2016rethinking}, and pre-train the separator with an L2 loss (until convergence) to help with training stability. We train $k$ DCGAN-like 
discriminators~\cite{radford2015unsupervised,stoller2018adversarial}, one per source:
\begin{gather}
    \text{L}^\text{real\_unpair}_{k}=\frac{1}{4} \bigg(D_k(\hat{Y}_{n,\omega,k}) - 0.9 \bigg)^{2} ,\notag \\
    \text{L}^\text{real\_pair}_{k}= \frac{1}{4}\bigg(D_k(Y_{n,\omega,k}) - 0.9 \bigg)^{2} ,\notag \\
    \text{L}^\text{fake\_unpair}_{k}=\frac{1}{4} \bigg(D_k(f(\hat{X}_{n,\omega})_k) - 0.1 \bigg)^{2} ,\notag \\
    \text{L}^\text{fake\_pair}_{k}= \frac{1}{4}\bigg( D_k(f(X_{n,\omega})_k) - 0.1 \bigg)^{2} ,\notag \\
    \text{L}_k=\text{L}^\text{real\_unpair}_{k} + \text{L}^\text{real\_pair}_{k} + \text{L}^\text{fake\_unpair}_{k} + \text{L}^\text{fake\_pair}_{k} .\notag
\end{gather}
With each iteration, we first update each discriminator $k$ (without updating the separator) based on the L$_k$ loss. Then, the L$_\text{adv}$ loss is used to train the separator (without updating the discriminator):
\begin{gather}
    \text{L}_\text{sep}=\frac{1}{K} \sum_{k} (D_k(f(X_{n,\omega})_k) - 0.9 )^{2} + (D_k(f(\hat{X}_{n,\omega})_k) - 0.9 )^{2} ,\notag \\
    \text{L}_\text{adv}=\text{L2}_\text{freq}+\gamma \text{L}_\text{sep} .\notag
\end{gather}

\vspace{-7mm}
\subsection{Deep feature loss}
\label{sec:deep_feature_losses}

Given a neural network $\phi$ with $J$ layers, the target $Y$ and estimated $\tilde{Y}$ magnitude spectrograms can be projected through $\phi$ to obtain their embeddings. Provided that $\phi$ is differentiable, the distance between target and estimated audio embeddings can be used to define deep feature losses at every layer $j$ of $\phi$~\cite{johnson2016perceptual}. The deep feature loss we use is based on the L2 distance~\cite{sahai2019spectrogram}: 
\begin{gather}
\text{L}_\text{feat}^{{\phi},j} = \frac{1}{C_{j}H_{j}W_{j}K}\sum_{c,h,w,k}| \phi_{j}(\tilde{Y}_{n,\omega,k}) -\phi_{j}(Y_{n,\omega,k})|^{2} , \notag
\end{gather}
where $C_{j}$$\times$$H_{j}$$\times$$W_{j}$ is the embedding's shape at each layer $j$. 
Deep feature losses were extended with a style reconstruction loss L$_\text{sty}^{\phi,j}$ capturing information on which features activate together~\cite{johnson2016perceptual,sahai2019spectrogram}. To that end, first, the gram matrix $G^{\phi(Y)}_{j}$ is defined as
\begin{gather}
G^{\phi(Y)}_{j} =  \frac{\psi_{Y} \psi^{\text{T}}_Y}{C_{j}H_{j}W_{j}} ,\notag
\end{gather}
where $\psi_{Y}$ corresponds to $Y$'s embeddings reshaped as $C_{j} \times H_{j}W{j}$. Accordingly, $G^{\phi(Y)}_{j}$ is of shape $C_j$$\times$$C_j$. Then:
\begin{gather}
\text{L}_\text{sty}^{\phi,j} = \sum_{c,c}
\left| G^{\phi(\tilde{Y})}_{j}-G^{\phi(Y)}_{j}\right|^{2} .\notag
\end{gather}
Our $\phi$ is a VGGish~\cite{hershey2017cnn} model pre-trained on AudioSet \cite{gemmeke2017audio}, and we use the above deep feature losses to regularize an L2$_\text{freq}$ loss~\cite{sahai2019spectrogram}:
\begin{gather}
 \text{L}_\text{deep\_feat} = \text{L2}_\text{freq} + \delta \text{L}_\text{feat}^{\phi,j} + \lambda \text{L}_\text{sty}^{\phi,j} \notag
\end{gather}

\section{Experimental setup}

\textbf{Task ---} We approach music source separation as defined by the MUSDB18 dataset~\cite{musdb18}, consisting of recovering 4 stereo stems (drums, bass, vocals and the residual `other') from stereo mixes. \mbox{MUSDB18 includes 150 songs (86 train, 14 val, 50 test) at 44.1kHz.} 

\vspace{1mm}
\noindent \textbf{Model ---} As noted above, we rely on OpenUnmix~\cite{stoter2019open} to consistently benchmark different training losses. 
Its original form is trained to independently separate vocals, bass, drums {\footnotesize\&} `other'. Hence, it trains a model per source. OpenUnmix later joins all these predictions via a multichannel Wiener post-filter {while} ensuring that all separations sum up to the mixture, further improving the separation~\cite{stoter2019open}. 
To simplify the pipeline, we modify OpenUnmix to predict all four sources using a single joint model: jOpenUnmix (Fig.~\ref{fig:UMXjoint}). In addition, we increase its latent from 512 feature maps to 1024. These minor modifications improve the results (trained with the original L2$_\text{freq}$ loss): vocals from 6.32\,dB to 6.4\,dB, drums from 5.73\,dB to 5.86\,dB,  {bass from 5.23\,dB to 5.28\,dB, and `other' from 4.02\,dB to 4.57\,dB.} jOpenUnmix keeps the original multichannel Wiener post-filtering.

\vspace{1mm}
\noindent \textbf{Losses setup ---}
For the adversarial loss, we set \mbox{$\gamma=0.5$~\cite{stoller2018adversarial}} and use an additional dataset where mixtures and isolated track recordings are unpaired.
We use 2065 additional unpaired mixtures from the Million Song Dataset~\cite{Bertin-Mahieux2011}. The unpaired vocals are 258 tracks from the iKala dataset~\cite{7178063} and 58~{vocals} tracks from the FreeSound Dataset~(FSD)~\cite{fonseca2017freesound}.
The unpaired drums and bass database consists of 114 and 83 tracks also from FSD, respectively. The 99 `other' tracks are from FSD---with the tags keys, brass, guitar, and strings.
Our DCGAN-like discriminators operate over stereo spectrograms and each consists of 9 CNN layers (of 4$\times$4 and stride of 2$\times$2 with 16, 32, 64, 128, 256, 512, 512, 512, 512 channels) and an MLP (with 3 layers of 1024, 32, 1 channels).
We set \mbox{$\beta = 0.05$} for L$_\text{dissim}$~\cite{huang2014deep}.
{We set \mbox{$\delta = \frac{1}{2}$} and \mbox{$\lambda = \frac{10}{3}$} for L$_\text{deep\_feat}$~\cite{sahai2019spectrogram}.
L$_{\text{MRS}}$ uses windows of 2048 and 1024 with 512 and 256 hop sizes. We use differentiable inverse-STFT layers for the losses requiring access to waveforms.

\vspace{1mm}
\noindent \textbf{Learning rates ---} 
Models are trained until convergence with Adam.
We reduce the learning rate $\times$0.3 after the validation loss does not improve for 80 epochs. It was key to independently adjust the learning rate for each loss.
Generally, $10^{-3}$ performed the best---except for $10^{-4}$ with L2$_\text{mask}$ loss, and $10^{-5}$ with L1$_\text{time}$ and L2$_\text{time}$ losses. Our discriminators and separator also use a learning rate of $10^{-3}$.

\begin{figure}[!t]
	\centering
	\includegraphics[trim={0cm 0cm 0cm 2.56cm},clip,width=0.9\columnwidth]{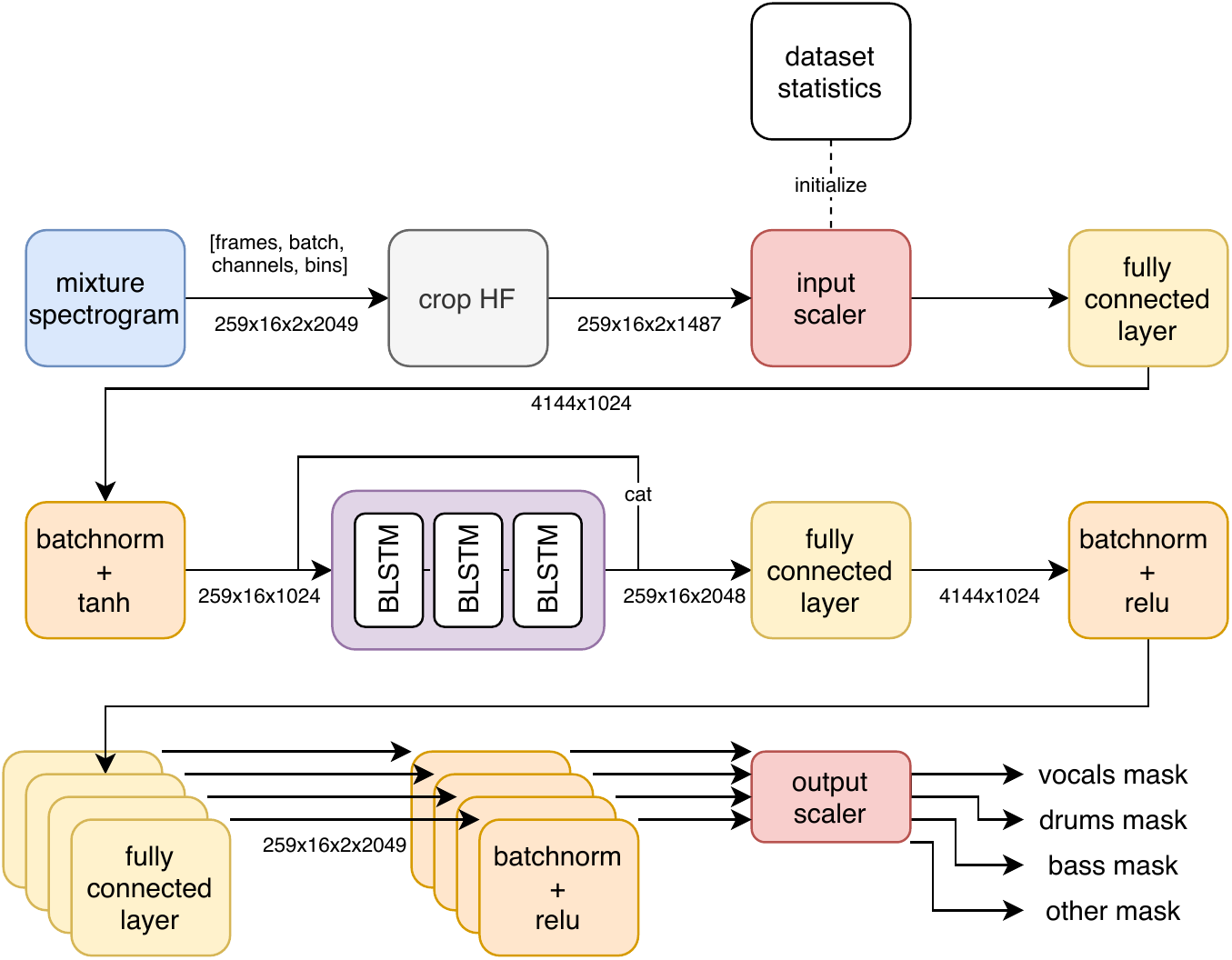}%
	\vspace{-2mm}
	\caption{jOpenUnmix model, the baseline.}
	\label{fig:UMXjoint}
	\vspace{-2mm}
\end{figure}

\section{OBJECTIVE EVALUATION}
We employ two sets of evaluation metrics: \textit{{i)}} the commonly used signal-to-distortion ratio (SDR, the higher the better) \cite{vincent2006performance,ward2018sisec}, in Table~1; and~\textit{{ii)}}~the~losses above used as evaluation metrics (the lower the better), in Fig.~\ref{fig:resultsglobal}.
In order to visually navigate the loss-based metrics in Fig.~\ref{fig:resultsglobal}, we standardize them to be in the same range.\footnote{Rows in Fig.~\ref{fig:resultsglobal} are standardized at zero mean and unit variance and clipped at $-1$ and $1$ for ease of visualization.} 
As a result, columns intuitively convey the performance of each loss across a wide range of loss-based metrics (vertical red lines denote bad performance, and blueish vertical lines denote good performance).
From Table~1 and Fig.~\ref{fig:resultsglobal}, we note that L1$_\text{mask}$ and L2$_\text{mask}$ perform among the worse, possibly because 
masks are ill-defined for the silent regions of the mixture. L1$_\text{time}$ and L2$_\text{time}$ also perform poorly, but other time-domain losses like SISDR$_\text{time}$, SDSDR$_\text{time}$, LOGL1$_\text{time}$ and LOGL2$_\text{time}$ perform well. We hypothesise that the logarithmic compression present in those losses helps attenuating low-energy artifacts, like the phase artifacts typically introduced by mask-based models. We also observe that the logarithmic compression might also help training, given that L1$_\text{time}$ and L2$_\text{time}$ metrics are surprisingly bad. 
This denotes that the model did not train as anticipated, since one expects a model trained on a specific loss to perform reasonably on a metric based on such loss.
We also note that SISDR$_\text{freq}$, LOGL1$_\text{freq}$ and LOGL2$_\text{freq}$ perform well.
Hence, scale invariant losses and their variants based on LOGL2 and LOGL1 are consistently delivering good results for both time-domain and spectrogram signals. 
The widely used L2$_\text{freq}$ loss also achieves good results, also when coupled with adversarial training and with the phase-sensitive loss (PSA). Yet, its separations did not improve when combined with deep feature losses. 
We studied combining the most promising losses (Combination: L2$_\text{freq}+$SISDR$_\text{freq}+$LOGL1$_\text{freq}$) with our regularizers (dissimilarity, adversarial and deep feature losses) without much success. 

\begin{table}[t!]
    \small
	\renewcommand{\arraystretch}{1.50}
	\label{obj_results}
	\centering
	\resizebox{0.8\columnwidth}{!}{\begin{tabular}{ l | c  c  c  c | c }
		\hline\hline
			 & Vocals & Drums & Bass & Other & Mean\\
		\hline
		L1$_\text{freq}$ & 5.95 & 5.58 & 4.24 & 3.90 & 4.92\\
		L2$_\text{freq}$ & 6.40 & 5.86 & 5.28 & \textbf{4.57} & \textbf{\color[HTML]{57bb8a}5.53}  \\
		L1$_\text{mask}$ & 5.10 & 4.14 & 2.88 & 2.53 & \textbf{\color{red}3.66} \\
		L2$_\text{mask}$ & 4.74 & 4.39 & 2.41 & 2.71 & \textbf{\color{red}3.56} \\
		L1$_\text{time}$ & 4.63 & 4.80 & 3.53 & 3.01 & \textbf{\color{red}3.99}\\
		L2$_\text{time}$ & 4.55 & 4.25 & 3.06 & 2.92 &\textbf{\color{red}3.70} \\
		SISDR$_\text{time}$ & 6.24 & 5.76 & 5.06 & 4.37 & 5.35 \\
		SISDR$_\text{freq}$ & 6.26 & 6.09 & \textbf{5.55} & 4.54&\textbf{\color[HTML]{57bb8a}5.61}\\
		SDSDR$_\text{time}$ & 6.02 & 5.84 & 4.99 & 4.37 & 5.30 \\
		LOGL1$_\text{time}$ & 5.95 & 5.82 & 5.23 & 4.35 & 5.34 \\
		LOGL2$_\text{time}$ & 5.88 & 5.81 & 5.04 & 4.10 & 5.21 \\
		LOGL1$_\text{freq}$ & 6.28 & 5.90 & 5.49 & 4.44 & \textbf{\color[HTML]{57bb8a}5.53}\\
		LOGL2$_\text{freq}$ & 6.15 & 5.79 & 5.38 & 4.36 & 5.42 \\
		PSA & 6.18 & \textbf{6.17}  & 5.10 & 4.33 & 5.44 \\
		Dissim$_\text{freq}$ & 6.04 & 5.66 & 5.17 & 4.38&5.31 \\
		MRS & 5.82 & 5.11 & 4.48 & 3.57& 4.80 \\
		DeepFeature & 6.14 & 5.89 & 4.80 & 4.30 & 5.28  \\
		Adversarial & \textbf{6.50} & 6.15 & 5.20 & 4.47 & \textbf{\color[HTML]{57bb8a}5.58} \\		
		Combination & 5.38 & 5.43 & 3.94 & 3.22 & 4.49 \\		
		\hline\hline
	\end{tabular}}
			\vspace{-1mm}
	\caption{Signal-to-distortion (SDR, in dB; the higher the better).}
		\vspace{-2mm}
\end{table}

\begin{figure}[!t]
	\centering
	\vspace{-2mm}
	\includegraphics[trim={2cm 0 1.1cm 0 0},clip,width=0.72\columnwidth]{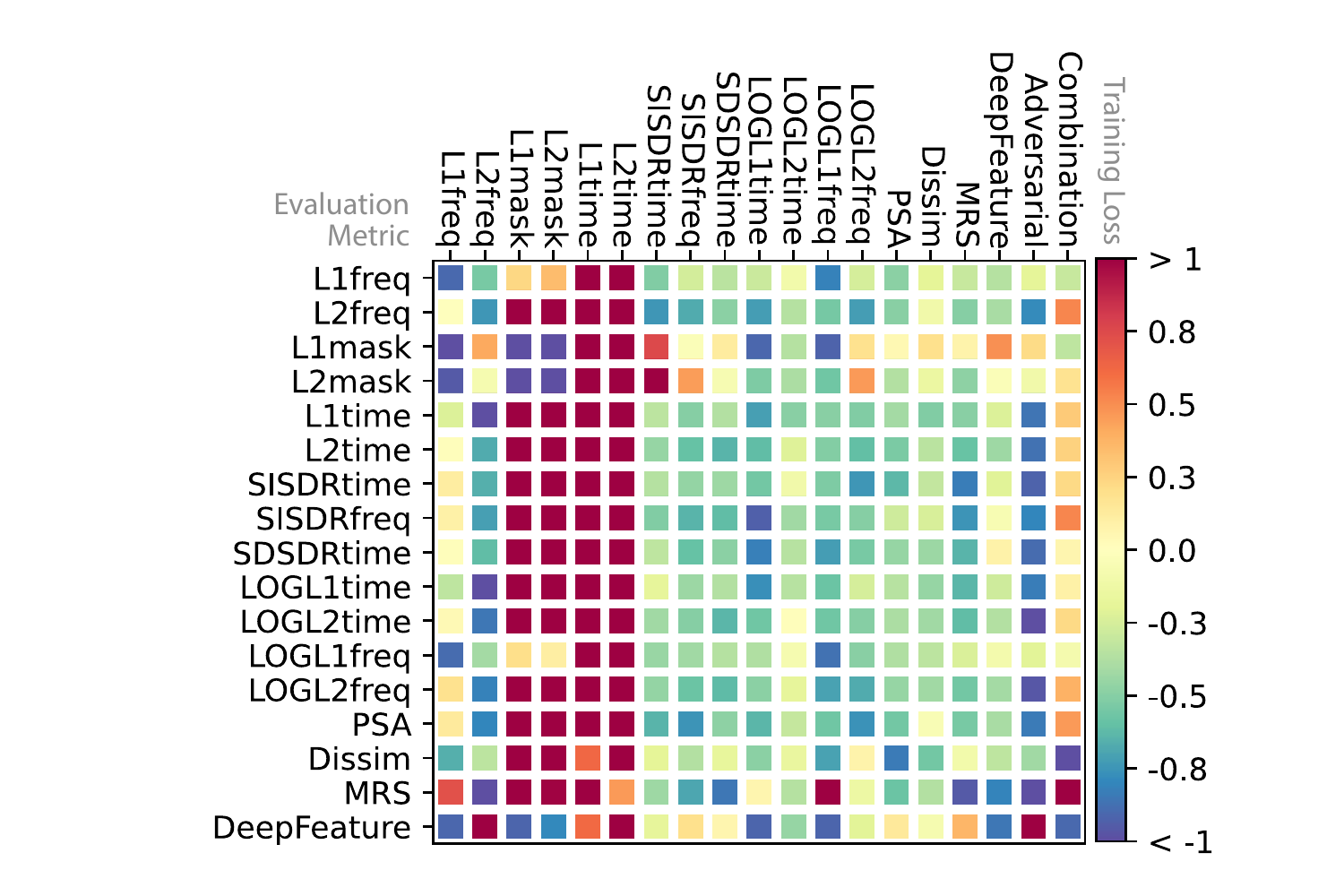}%
		\vspace{-4mm}
	\caption{Loss-based metrics$^2$ over the test set (the lower the better).}
	\label{fig:resultsglobal}
		\vspace{-2mm}
\end{figure}

\section{Subjective Evaluation}

Since running a subjective evaluation for all losses is barely practicable, we relied on the results above and informal listening to select the five best performing ones: LOGL1$_\text{freq}$, LOGL1$_\text{time}$, Adversarial, L2$_\text{freq}$ and SISDR$_\text{freq}$.
For each loss, we evaluated vocals, bass, drums and `other' separations for 4 different songs. 
15 participants were asked to rate the global quality of the separations using a 0--100 scale~\cite{schoeffler2018webmushra}. 
The mean opinion scores (MOS) are reported in Table 2. We find that LOGL1$_\text{freq}$ and L2$_\text{freq}$ perform similarly (t-test: $p=0.94$) and outperform the rest (t-test: $p<0.05$) except SISDR$_\text{freq}$ (t-test: $p<0.27$).
Yet, overall, all separations lie in the upper \textit{fair} (40--60) range. Notably, recent music source separation studies also reported MOS scores around the \textit{fair} (40--60) and \textit{good} (60--80) range~\cite{defossez2019music}. Hence, despite recent progress, 
the task is far from being solved. 
Importantly, if we compare the results of this subjective test with the evaluation metrics in section 4, we see that it exists a $\approx$\,0.1\,dB~SDR difference between L2$_\text{freq}$ (best human rating) and SISDR$_\text{freq}$ (best SDR score). This discrepancy is remarkable, since a $\approx$\,0.1\,dB improvement is often enough to claim state-of-the-art results in the literature~\cite{defossez2019music,li2021sams,takahashi2020d3net}---although L2$_\text{freq}$ (with lower SDR) seems to be perceptually preferable.
Hence, does it exist a metric that correlates better with human judgment than SDR?
To answer this question, we cross-correlate our loss-based metrics (Fig.~\ref{fig:resultsglobal}, rows) with the results of the subjective test (Table~2, rows). This experiment (Fig.~\ref{fig:corrMOS}) 
shows that spectrogram-based metrics, like L1$_\text{freq}$ or SISDR$_\text{freq}$, are the ones correlating best with human judgment.






\begin{table}[t!]
    \small
	\renewcommand{\arraystretch}{1.50}
	\label{taresults2}
	\centering
	\resizebox{\columnwidth}{!}{\begin{tabular}{  l | c  c  c  c  c }
	\hline\hline
		 & L2$_\text{freq}$ & SISDR$_\text{freq}$ & LOGL1$_\text{time}$ & LOGL1$_\text{freq}$ & Adv \\
		\hline
		Vocals & 57.24$_{\pm17.9}$ & 53.33$_{\pm17.0}$ & 54.83$_{\pm19.1}$ & \textbf{59.10}$_{\pm17.0}$ & 54.42$_{\pm18.4}$ \\
		 Drums & \textbf{60.58}$_{\pm15.6}$ & 59.60$_{\pm18.5}$ & 59.10$_{\pm19.8}$ & 59.76$_{\pm21.7}$ & 54.40$_{\pm18.2}$\\
		 Bass & 61.62$_{\pm20.3}$ & 62.80$_{\pm19.9}$ & 57.19$_{\pm23.2}$ & 62.13$_{\pm22.5}$ & \textbf{64.28}$_{\pm21.0}$  \\
		 Other & \textbf{56.90}$_{\pm18.6}$ & 52.22$_{\pm19.0}$ & 48.98$_{\pm18.5}$ & 54.88$_{\pm18.5}$ & 48.37$_{\pm20.4}$  \\
		\hline 
		 Mean & \textbf{59.09}$_{\pm18.2}$ & 56.99$_{\pm19.0}$ & 55.01$_{\pm20.5}$ & 58.96$_{\pm20.1}$ & 55.39$_{\pm20.2}$  \\
	\hline\hline
	\end{tabular}}
	\vspace{-3mm}
	\caption{MOS ($\pm$ standard deviation) subjective test.}
	\vspace{-1mm}
\end{table}

\begin{figure}[!t]
	\centering
	\includegraphics[trim={1cm 0cm 0cm 10cm 0},clip,width=\columnwidth]{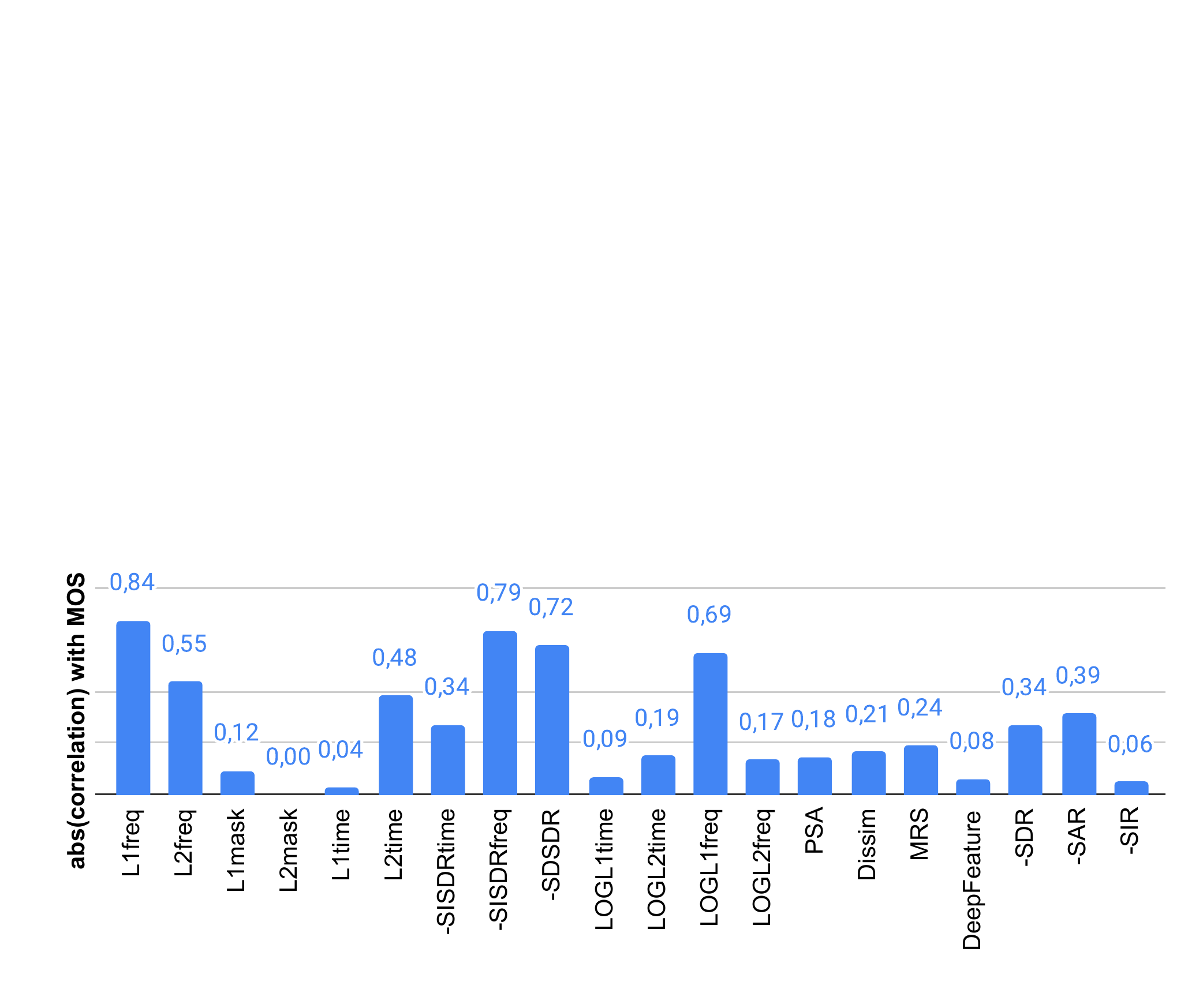}%
	\vspace{-7mm}
	\caption{Correlation between the loss-based metrics (Fig.~\ref{fig:resultsglobal}, rows) and the MOS human ratings (Table~2, rows).}
	\label{fig:corrMOS}
	\vspace{-1mm}
\end{figure}

\section{Discussion}

We extensively reviewed the most representative audio source separation losses and benchmarked those for the task of music source separation. After evaluating those objectively and subjectively, we recommend training with the following spectrogram-based losses: L2$_\text{freq}$, SISDR$_\text{freq}$, LOGL2$_\text{freq}$ or LOGL1$_\text{freq}$ with, potentially, phase-sensitive objectives and adversarial regularizers. 
We also found that LOGL1$_\text{time}$ can deliver competent results, and we do not recommend using mask-based losses. 
That said, we want to emphasize the limitations of our experimental setup. Since our experiments rely on OpenUnmix, a spectrogram-based model predicting mask filters, our conclusions might be constrained to this family of models. For example, while L1$_\text{time}$ and L2$_\text{time}$ performed poorly in our setup, these are widely used to train waveform-based models~\cite{defossez2019music,pons2020upsampling,stoller2018wave}.
Finally, we also cross-correlated the set of metrics we investigated with the results of our subjective test. Out of this experiment,
we conclude that it could be informative if future works on music source separation also reported spectral distortion metrics (like L1$_\text{freq}$) together with SDR-based ones. Note that such metrics are already being used to evaluate speech synthesis models~\cite{kubichek1993mel}.

\bibliographystyle{IEEEbib}
\bibliography{ISMIRtemplate}

\end{document}